\documentstyle[12pt]{article}

\def\beq{\begin{equation}}
\def\eeq{\end{equation}}
\def\bea{\begin{eqnarray}}
\def\eea{\end{eqnarray}}
\begin{document}
\topmargin 0pt
\oddsidemargin=-0.4truecm
\evensidemargin=-0.4truecm
\renewcommand{\thefootnote}{\fnsymbol{footnote}}
\newpage
\setcounter{page}{0}
\begin{titlepage}
\begin{flushright}
FTUV/94-27\\
IFIC/94-25
\end{flushright}
\begin{center}
{\large A SIMPLE WAY TO ESTIMATE THE VALUE OF
$\bar{\alpha}\equiv\alpha(m^2_Z)$}
\vglue 0.5cm
{\large R.B.Nevzorov, A.V.Novikov}\\
{\em ITEP, Moscow 117259 Russia}\\
\vspace{0.3cm}
{\em and}\\
\vspace{0.3cm}
{\large M.I.Vysotsky
\footnote{Permanent address: ITEP, Moscow 117259, Russia}}\\
{\em Instituto de Fisica Corpuscular (IFIC--CSIC)\\
Departamento de Fisica Teorica, Universitat de Valencia\\
Dr. Moliner 50, 46100 Burjassot (Valencia), Spain}\\
\vspace{0.5cm}
\end{center}
\begin{abstract}
To obtain the value of electromagnetic coupling constant at $q^2=m^2_Z$,
$\bar{\alpha}$, which plays a key role in electroweak physics one has
to integrate the
cross-section of $e^+e^-$-annihilation into hadrons divided by $(s-m^2_Z)$
over $s$ from threshold to infinity. By combining, for each flavor channel,
the contribution of lowest resonance with the perturbative QCD continuum,
we obtain $1/\bar{\alpha} = 128.90\pm 0.06 $  a result which is close to
known
result obtained with purely experimental inputs, i.e. $1/\bar{\alpha} =
128.87\pm 0.12$.
\end{abstract}
\end{titlepage}
\vspace{2cm}
\vspace{.5cm}
\renewcommand{\thefootnote}{\arabic{footnote}}
\setcounter{footnote}{0}
\newpage
The detailed analysis of electroweak observables starts from three input
parameters: $G_{\mu}$, the Fermi coupling constant
(extracted from muon decay), $m_Z$, the $Z$-boson mass (measured at LEP) and
$\bar{\alpha}$, the electromagnetic coupling constant at $q^2 =m^2_Z$, obtained
from dispersion relations.
In fact, a Born aproximation to the minimal standard model which starts with
$\bar{\alpha}$ (rather than $\alpha\equiv\alpha(0)=1/137.0359895(61)$ )
reproduces the precise experimental values of the $Z$-decay parameters
(obtained at LEP) and of the $W$ mass (obtained at
hadron colliders) with unexpectedly high accuracy \cite{1}, \cite{2}.
For example for the ratio of vector and axial coupling
constants of the $Z$-boson to charged leptons one obtains in this
$\bar{\alpha}$ Born approximation \cite{2}:

\beq
[g_V/g_A]_{\bar{\alpha}}=0.0753(12)~,
\label{1}
\eeq
while the latest experimental numbers are \cite{3}:
\beq
[g_V/g_A]_{LEP}=0.0711(20)~,
\label{2}
\eeq
\beq
[g_V/g_A]_{LEP+SLD}=0.0737(18)~.
\label{3}
\eeq
If instead of $\bar{\alpha}$ one uses $\alpha(0)$, then one gets:
$$
[g_V/g_A]_{\alpha}=0.152 \,\, ,
$$
which is about $40\sigma's$ away from experiment as was stressed in
\cite{2}.
The value of $\bar{\alpha}$ is of fundamental importance, and its error
determines the uncertainty
in the theoretical prediction (\ref{1}).

$\bar{\alpha}$ is defined through the following formulas:
\beq
\bar{\alpha}=\frac{\alpha}{1-\delta\alpha} \;\; ,
\label{4}
\eeq
\beq
\delta\alpha=\Sigma'_{\gamma}(0) -\frac{\Sigma_{\gamma}(m^2_Z)}{m^2_Z} \;\; ,
\label{5}
\eeq
where in (\ref{5}) charged leptons and five quark flavor
contributions in photon polarization operator should be taken into account.
Contributions of $(t\bar{t})$ and $(W\bar{W})$ loops may be omitted in
(\ref{5}); they are numerically small and usually are attributed to proper
electroweak radiative corrections \cite{4}. The following integral
representation for $\delta\alpha$ is valid:
\beq
\delta\alpha=\frac{m_Z^2}{4\pi^2\alpha}\int\frac{\sigma_{e^+e^-\to \mbox{\rm
all}}(s)} {m_Z^2 -s}ds \;\; ,
\label{6}
\eeq
where integral goes from threshold to infinity and its principal value at
$s=m_Z^2$ should be taken. The lepton contribution of $e$, $\mu$, and $\tau$
to
(\ref{6}) are readily calculated and one gets:  \beq
(\delta\alpha)_l=\frac{\alpha}{3\pi}[\Sigma\ln\frac{m_Z^2}{m_l^2}-\frac{5}{3}]=
\frac{\alpha}{3\pi}[22.5+11.8+6.2]=0.0314~.
\label{7}
\eeq
For the hadronic contribution in  \cite{5} the following number was obtained
(see
also \cite{6}):
\beq
(\delta\alpha)_h=0.0282(9)~.
\label{8}
\eeq
To obtain this number the experimental cross-section for
$e^+e^-$-annihilation into hadrons below $s_0 =(40{\rm GeV})^2$ and parton
model result above $s_0$ was used in \cite{5} and \cite{6}.

The difficulty in the theoretical determination of $(\delta\alpha)_h$
comes from its logarithmic dependence on the infrared cutoff.
As it was mentioned in \cite{7}, the result of the dispersion
calculation of $(\delta\alpha)_h$ can be reproduced by using perturbative
QCD with the following effective "quark masses":
\bea
m_u=53 {\rm MeV} \; , \;\; m_d =71 {\rm MeV} \; , \;\; m_s = 174 {\rm MeV}
\; , \nonumber \\
m_c = 1.5 {\rm GeV} \; , \;\; m_b = 4.5 {\rm GeV}~.
\label{9}
\eea
Unfortunately one can not attribute any physical meaning to these values of
$m_u$ and  $m_d$.

Our aim here is to present simplest sensible model for $\sigma_{e^+e^- \to
{\rm hadrons}}$ which can simulate the result given in (\ref{8}). To do this
we use one physical resonance ($\rho, \omega, \varphi,
J/\Psi$ and $\Upsilon$) at the beginning of spectrum and then starting from
$E_i = m_i +\frac{\Gamma_i}{2}$ the QCD improved parton model continuum
in each quark channel.

For resonance contribution we use Breit-Wigner formula:
\beq
\sigma_{ee} = \frac{3\pi \Gamma_{ee}\Gamma}{E^2[(E-m)^2+\Gamma^2/4]}~.
\label{10}
\eeq
Substituting
in (\ref{6}), neglecting terms of the order of $(m/m_Z)^2$ and integrating
from $-\infty$ to $m+\frac{\Gamma}{2}$ we obtain:
\beq
(\delta\alpha)_{{\rm resonance}}
= \frac{3}{\alpha}\frac{\Gamma_{ee}}{m}\frac{3}{4}~.
\label{11}
\eeq
Thus vector meson contributions into $\delta\alpha$ are:
\bea
\begin{tabular}{lccccc}
 & $\rho$ & $\omega$ & $\varphi$ & $J/\Psi$ & $\Upsilon$ \\
 $\delta\alpha$ & 0.00274(13) & 0.00024 & 0.00042 & 0.00053 & 0.000045~
,\\
\end{tabular}
\label{12}
\eea
where we take into account experimental
uncertainty for $\rho$-meson contribution as the only noticeable.

For continuum contribution we use the following formulas:
\beq
\sigma_{I=1}=2\pi\frac{\alpha^2}{s}(1+\frac{\alpha_s(s)}{\pi})~,
\label{13}
\eeq
\beq
\sigma_{I=0}=\frac{2\pi}{9}\frac{\alpha^2}{s}(1+\frac{\alpha_s(s)}{\pi})~,
\label{14}
\eeq
\beq
\sigma_{s\bar{s}}=\frac{4\pi}{9}\frac{\alpha^2}{s}(1+\frac{\alpha_s(s)}{\pi})~,
\label{15}
\eeq
\beq
\sigma_{c\bar{c}}=\frac{16\pi}{9}\frac{\alpha^2}{s}\sqrt{1-\frac{4m_c^2}{s}}
(1+\frac{2m_c^2}{s})(1+\frac{\alpha_s(s)}{\pi})~,
\label{16}
\eeq
\beq
\sigma_{b\bar{b}}=\frac{4\pi}{9}\frac{\alpha^2}{s}\sqrt{1-\frac{4m^2_b}{s}}
(1+\frac{2m_b^2}{s})(1+\frac{\alpha_s(s)}{\pi})~,
\label{17}
\eeq

where we use for $\alpha_s(s)$ the following formula:
\beq
\alpha_s(s)=\frac{12\pi}{(33-2n_f)\ln s/\Lambda^{(n_f)^2}}
\label{18}
\eeq
with $\alpha_s(m_Z)=0.129(5)$ as an input (this one loop value corresponds
to 0.125(5) at three loops, which is extracted from latest LEP data \cite{2}).
We take $n_f=5$ for
$s>m_\Upsilon^2$, $n_f=4$ for $m_\Upsilon^2>s>m_{J/\Psi}^2$, $n_f=3$
for $m_{J/\Psi}^2>s>m_\varphi^2$ and $n_f=2$ for $m_\varphi^2>s>(m_\rho
+\Gamma_\rho/2)^2$. This corresponds to $\Lambda^{(5)}=160MeV,
\Lambda^{(4)}=220MeV,
\Lambda^{(3)}=270MeV$ and $\Lambda^{(2)}=300MeV$.

Substituting (\ref{13}) -- (\ref{18}) into (\ref{6}) with $m_c = m_b =0$ we
get:
\bea
\begin{tabular}{lccccc}
             &  $I=1$  &  $I=0$  & $s\bar{s}$  & $c\bar{c}$  & $b\bar{b}$ \\
$\delta\alpha$ & 0.01174 &0.00133 & 0.00249   & 0.00741   & 0.00123
\end{tabular}
\label{19}
\eea

Summing up contributions of (\ref{12}) and (\ref{19}) we get:
\beq
(\delta\alpha)_h=0.0282~ , \;\; \bar{\alpha} =(128.87)^{-1}
\label{20}
\eeq
Comparing with obtained by integrating experimental data results (\ref{8})
$(\delta\alpha)_h=0.0282$
and $\bar{\alpha}=[128.87(12)]^{-1}$ we see that agreement is astonishing.
The contribution of $\alpha_s$ correction in (\ref{19}) is rather small,
$0.00087 + 0.00010 + 0.00018 + 0.00042 + 0.00006 = 0.00163$, so even
if light gluino octet slow down $\alpha_s$ running in order to accomodate
$\alpha_s$ values measured at quarkonium decays \cite{100} $(\delta\alpha)_h$
will decrease by 0.0002 only.

We have to make a few comments:

(1) taking contributions
$\sim \alpha_s^2$  in continuum cross-section into account
and using next-to-leading order formula for $\alpha_s(s)$ we increase
$(\delta\alpha)_h$ by 0.00045; negative contribution of $\sim \alpha_s^3$
term appeares to be approximately two times larger. In beauty and charm
channels third loop gives much smaller contribution than second (numerically
both are negligible) so we can trust our continuum calculation. In strange
channel third loop contribution equals that of second, while in $I=1$ and
$I=0$ channels it is
two times larger. So below, say, 1.5GeV perturbative continuum can not be
approved. Allowing physical continuum variation at the level of
$\pm15\%$ around tree plus
one loop perturbative continuum value in the domain 1 - 2GeV we get
$\pm 0.0004$ variation
in $(\delta\alpha)_h$;

(2) experimental uncertainty in $\Gamma_{ll}$
of vector resonances lead to $(\delta\alpha)_h$ variation of the order of
0.0002, while that in $\alpha_s(m_Z)$ -- to 0.0001 variation. Both are
small compared with the uncertainty 0.0009 in (\ref{8});

(3)subtracting from the $\rho$ contribution the integral over Breit-Wigner
formula from $-\infty$ to two pion threshold we diminish it by:
\beq
\delta\alpha_{sub}=\frac{3\Gamma_{ee}}{2\pi\alpha m_{\rho}}2
\arctan \Gamma_{\rho}/(2(m_{\rho}-2m_{\pi}))=0.00017;
\label{101}
\eeq

(4) taking into
account heavy quark masses $m_c =
1.6$ GeV, $m_b = 4.7$ GeV, we
decrease $(\delta\alpha)_h$ correspondingly by:
\beq
(\delta\alpha_h)_m= 0.00031 + 0.00008 = 0.00039;
\label{21}
\eeq

(5) finally, at energies $E=m_i +\frac{\Gamma_i}{2}$ our model curve for
$\sigma_{e^+e^- \to {\rm hadrons}}$  is
discontinuous. To understand $(\delta\alpha)_h$ sensitivity for the details
of the model we change it in the following way: we continue $\rho$,
$\omega$, $\varphi$ and $J/\Psi$ resonance  curves up to their intersection
with quarks continuum. In this way $(\delta\alpha)_h$ increases :
\beq
\delta(\delta\alpha)_h=0.00039
\label{22}
\eeq

Subtracting from (\ref{22}) sum of (\ref{21}) and (\ref{101})
and taking uncertainty from point (1) above for total shift we get:
\beq
(\delta\alpha)_h=0.0280(4), \ \   \bar{\alpha} = (128.90(6))^{-1}
\label{23}
\eeq
So it is evident that the value of $(\delta\alpha)_h$ is rather insensitive to
the details of the model of $\sigma_{e^+e^- \to {\rm hadrons}}$. More
refined model which takes all known resonances in each flavour channel
into account gives $(\delta\alpha)_h=0.0275(2)$
\cite{8}.

For real progress in diminishing error in (\ref{8}) systematic error in
cross section of $e^+e^-$-annihilation into hadrons in
background region below 3 GeV should be improved \cite{5}, \cite{9}.

We thank  B.V.Geshkenbein, A.I.Golutvin, V.L.Morgunov, V.A.Novikov, L.B.Okun,
V.L.Telegdi
and J.W.F.Valle for discussions, and F.Jegerlehner for correspondence
concerning his results. M.V. is grateful to Instituto de
Fisica Corpuscular - C.S.I.C. and Departament de Fisica
Teorica, Universitat de Valencia,
where this work was finished, for warm hospitality.
This work was supported in part by the ISF grant MRW000.

\end{document}